\documentclass[10pt,letterpaper,twocolumn]{article} 

\usepackage{ol2}
\usepackage{amsmath,amssymb,graphicx,subfigure,color}
 \newcommand{\enquote}[1]{``#1''}

\begin{document}

\twocolumn[
\title{Self-synchronizing scheme for high speed computational ghost imaging}

\author{Jinli Suo,$^1$ Yudong Xiao,$^2$ Liheng Bian,$^1$ Lei Zhang,$^2$ and Qionghai Dai,$^{1,*}$}

\address{
$^1$Department of Automation, Tsinghua University, Beijing 100084, China\\
$^2$Graduate School at Shenzhen, Tsinghua University,  Shenzhen 518055, China\\
$^*$Corresponding author: qhdai@tsinghua.edu.cn
}

\begin{abstract}Computational ghost imaging needs to acquire a large number of correlated measurements between reference patterns and the scene for reconstruction, so extremely high acquisition speed is crucial for fast ghost imaging.
With the development of technologies, high frequency illumination and detectors are both available, but their synchronization needs technique demanding customization and lacks flexibility for different setup configurations.
This letter proposes a self-synchronization scheme that can eliminate this difficulty by introducing a high precision synchronization technique and corresponding algorithm.
We physically implement the proposed scheme using a 20kHz spatial light modulator to generate random binary patterns together with a 100 times faster photodiode for high speed ghost imaging, and the acquisition frequency is around 14 times faster than that of state-of-the-arts.\end{abstract}
\ocis{110.0110, 110.1758, 110.3010.}
]

\noindent Ghost imaging that takes root from  quantum optics\cite{pittman1995optical}\cite{strekalov1995observation} and then extends to classical light field \cite{bennink2002two}\cite{ferri2005high}, can recover the scene information using its correlated measurements with a reference beam.  Computational imaging \cite{shapiro2008computational} is later proposed to reconstruct the scene using a computer controlled pattern sequence instead of physical reference arm. The relationship and comparison among above three variants---quantum, classical and computational---are comprehensively reviewed by Erkmen and  Shapiro in \cite{erkmen2010ghost}. Such an imaging technique without a spatially resolvable sensor holds great potential for building compact optical harsh systems and makes good use the high performance of the single-pixel bucket photo-detectors. Large progresses in ghost imaging have been made in the past years. Recently, Sun el al.\cite{sun20133d} combine ghost imaging with computer vision algorithms (photometric stereo) and successfully conduct 3D reconstruction using multiple single pixel detectors and one projector. Later, Welsh et al.\cite{welsh2013fast} extend the gray scale computational ghost image to full-color using three detectors that response to different wavelengths.

Among all of the above state-of-the-arts, each computational ghosting imaging approach needs to collect a large number of single-pixel measurements for the final imaging \cite{erkmen2009signal}\cite{meyers2008ghost}, usually by a pair of illumination and detector synchronized either manually or automatically.
Although one can utilize the redundancy in nature images \cite{welsh2013fast}\cite{katz2009compressive}\cite{compressive_scientificreport} or a coarse-to-fine scheme \cite{multiscale_ao}\cite{adaptive_wavelet_oe} to reduce the number of measurements, the required number of measurement is still quite large, especially in the real experiments where orders of magnitude times more measurements are needed to compensate the system noise and influences from external factors. As far as we know, by utilizing the mechanisms of DLP smartly \cite{sun20133d}\cite{welsh2013fast} state-of-the-art fast computational ghost imaging is able to capture no higher than 1.5K correlated measurements per second. However, it is still quite slow since hundreds of thousands of measurements are necessary for visually pleasant reconstruction. Therefore, ghost imaging is really time consuming and this drawback largely limits its real applications.

\begin{figure}[t]
  \centering
  \includegraphics[width=0.94\linewidth,height=0.54\linewidth]{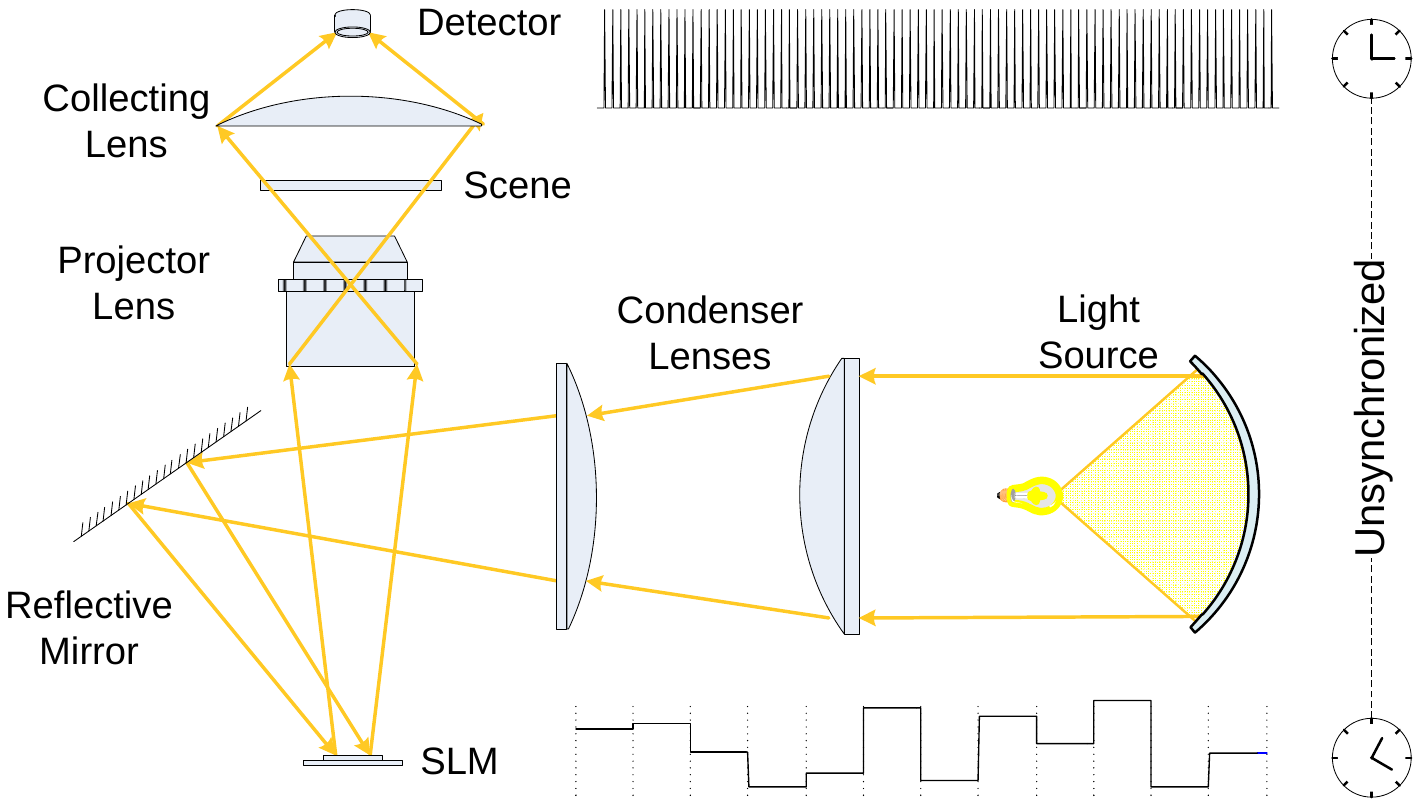}
  \caption{The schematic setup of the proposed fast ghost imaging system. The high speed photodiode (detector) and the SLM (illumination) both work at fixed frequencies but are not synchronized, while the projected patterns from SLM and the acquired measurements by detector are self synchronized by an algorithm.}\label{fig:lightpath}
  \vspace{-3mm}
\end{figure}

Raising frequency of illumination and detector can increase the acquisition frequency naturally\cite{faccio2013ghost}, but the synchronization becomes more and more technically demanding with the increase of this frequency, because hardware expertise (in both illumination and detector) is necessary for the synchronization. In addition, the close bounding between the synchronization and hardware makes the system less flexible. Therefore, we attempt to propose a general and flexible fast ghost imaging scheme being able to readily incorporating currently available high speed illuminations and detectors, which both work at constant frequencies and are synchronized computationally, as illustrated in Fig.~\ref{fig:lightpath}.

This letter draws some inspirations from the high-speed hyper-spectral imaging method proposed by Han et al.\cite{han2011fast}, who use a "white board" with known reflectance behaviours to synchronize illumination and camera working at constant frequencies and capture the scene's reflectance fast. However, introducing such an auxiliary object is inapplicable for our task since its information is coupled with that of scene. To address this problem, we introduce designed patterns with the corresponding measurements largely different from that of the random patterns\cite{welsh2013fast}, to synchronize the illumination and detector. Then, we can exclude these synchronization patterns and reconstruct the scene using off-the-shelf ghost imaging algorithms.

\begin{figure}[t]
  \centering
  \includegraphics[width=0.98\linewidth]{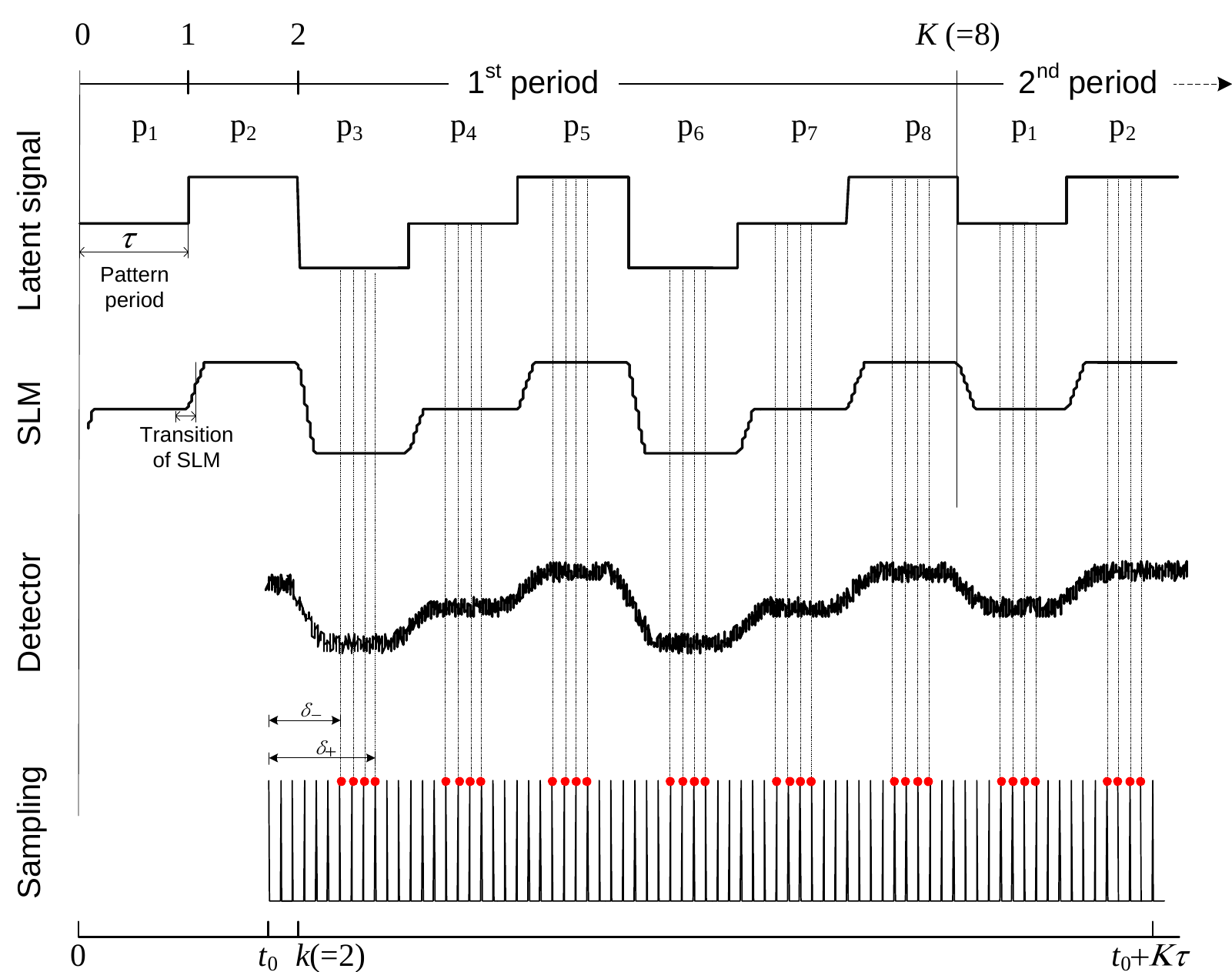}
  \caption{The illustration of synchronization between high speed cyclic pattern sequence and detector's measurements. The unsynchronization includes both integral and fractional components, denoted as $k$ and $\{\delta_-,\delta_+\}$.}\label{fig:align}
  \vspace{-2mm}
\end{figure}

Currently, we can use a programmable spatial light modulator (SLM), e.g. digital micro-mirror device (DMD), to shed binary coded illuminations fast with a small elapse during the 0/1 transition, which is determined by the transition and stabilization velocity of SLM units. This elapse is quite small but non-ignorable for high speed illumination of thousands Hz. Therefore, we can regard the latent measurement for ghost imaging as a close-to-ideal square wave, as illustrated in the top two rows of Fig.~\ref{fig:align}. Note that here we assume a gradual transition between adjacent illumination patterns in general, there may be some other transition ways. For example, there is a reset operation of micro mirror position in DMD modulation, so a zero measurement will occur in the transition slot. On the other hand, photodiodes with a much higher speed (up to GHz) are widely available and can be sampled at lower frequencies using an acquisition card. Therefore, we can conduct data acquisition multiple or even orders of magnitudes faster than illumination patterns and combine with an algorithm to avoid the influences from inter-pattern transition and analog-to-digital noise, as illustrated in the third row of Fig.~\ref{fig:align}.

Without loss of generality, we assume both the SLM and detector work at constant frequencies denoted as  $f_{L}$ and $f_{D}$ respectively, and the latter is integral times of the former $f_{D}=f_{L}\times n$ (e.g., $n=100$). Mathematically self-synchronization turns into computing the optimum sampling points, as illustrated by the red points in the bottom row of Fig.~\ref{fig:align},  and their corresponding illumination patterns.
To synchronize these two modules, we keep one working repeatedly and start the other from an arbitrary time instant $t_0$, then the misalignment can be described by two parameters: the integral misalignment $k \in \{1, 2,\cdots, K\}$ with $K$ being the total number of illumination patterns, i.e., index of the starting illumination pattern with corresponding measurement captured; the fractional misalignment 
described as a reliable index set $[\delta_-,\delta_+]\in\{1,2,\cdots,n\}$, whose size $|\delta_+-\delta_-|$ is determined by the time span of SLM's stabilized stage.


\begin{figure}[h]
  \centering
  \includegraphics[width=0.96\linewidth]{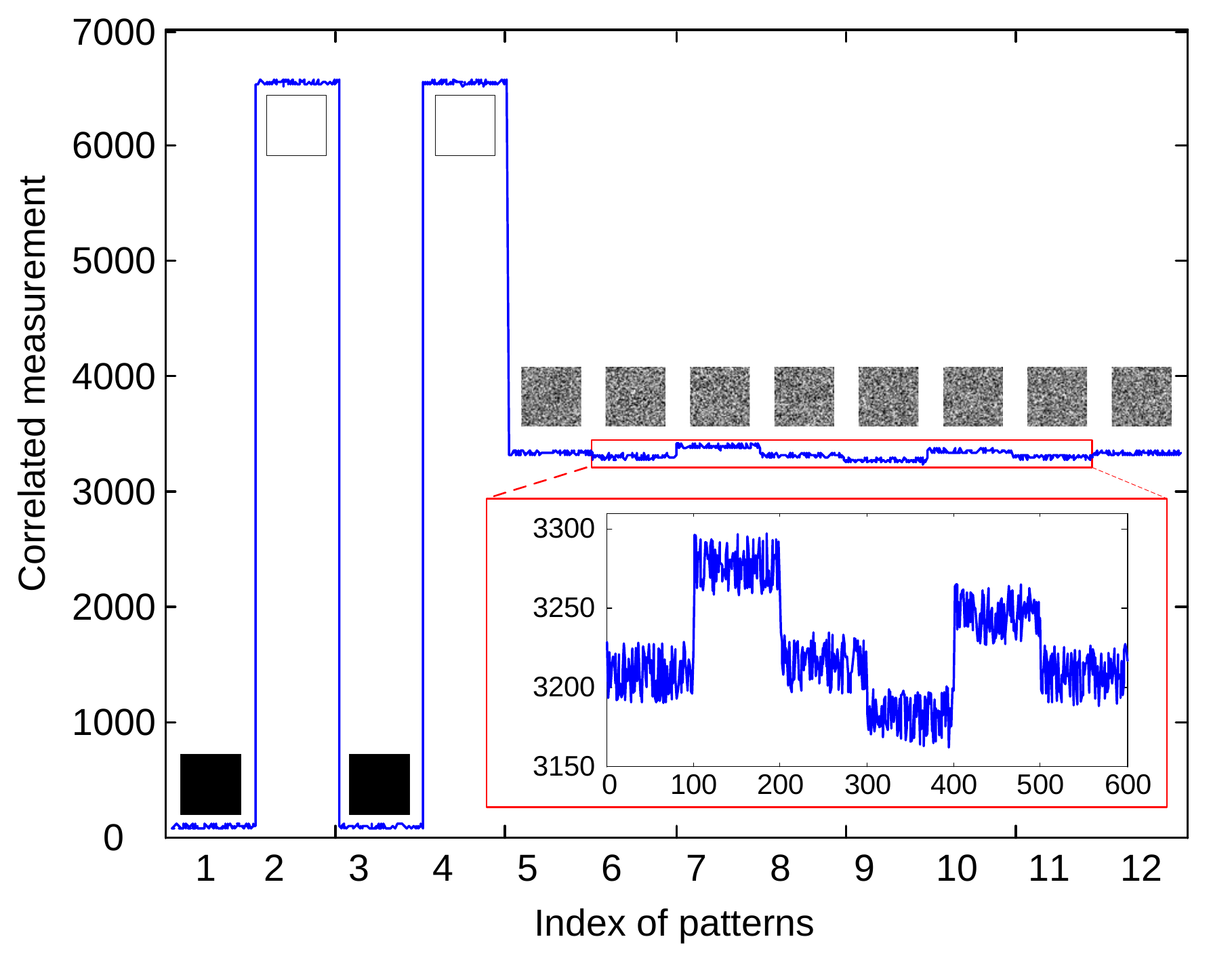}
  \vspace{-2mm}
  \caption{The illustration of fractional alignment, with two groups of synchronization patterns and more binary random patterns for encoding 2D scene information. The closeup view of the latter part is displayed in the bottom right region and highlighted with red outlines.}\label{fig:coarse}
\end{figure}

The misalignment parameters $k$ and $\{\delta_-,\delta_+\}$ between illumination patterns $\{{\bf p}_i;~i=1,\cdots, K\}$ and the measurements $\{m_i;~i=1,\cdots, nK\}$ can be estimated easily by prefixing some illumination patterns whose measurements are largely different from those of random patterns, as shown in Fig.~\ref{fig:coarse}. The adopted alternate all-0 and all-1 synchronization patterns are somewhat similar to but of a different usage from that in \cite{welsh2013fast}.
Assuming $L$ groups of synchronization patterns are prefixed, the optimum misalignment parameters are estimated by minimizing the measurement difference within the all-0 or all-1 pattern set (numerator term in Eq.~\ref{eq:syn}) and maximizing that between measurement sets corresponding to all-0 and all-1 illumination (denominator term in Eq.~\ref{eq:syn})
\begin{equation}
(k^*, \delta^*)\!=\!\arg\min\frac{\sum_{l=1,\!\cdots\!,2L}\left<|\Omega_{k,l}-\left<\Omega_{k,l}\right>|\right>}
{\left|\sum_{l=1,\!\cdots\!,L}\!\left<\Omega_{k,2l-1}\!\right>\!-\!\sum_{l=1,\!\cdots\!,L}\!\left<\Omega_{k,2l}\!\right>\right|}\label{eq:syn}
\end{equation}
in which
\begin{equation}
\Omega_{k,l}\!=\!\{m_{(K-k+l-1)\%K\cdot n+\delta_-},\!\cdots\!,\!m_{(K-k+l-1)\%K\cdot n+\delta+}\}.
\end{equation}
Here $\Omega$ represents a set of measurements in the stabilized stage of an illumination pattern, $\%$ denotes the remainder operator, and $\left<\cdot\right>$ denotes average operation along temporal dimension. For fast computation, we can firstly locate the coarse position of the synchronization sequence according to their quite large or small measurements, and then optimize the parameters via traversing searching locally. The candidate parameter is no larger than 2$n$, so the algorithm is quite efficient.
Experimentally, only 2-3 such alternative pattern groups are sufficient for synchronization.

After estimating $k$ and $\{\delta_-,\delta_+\}$, we then cycle shift the pattern sequence $\{{\bf p}_i\}$ to get
\begin{equation}
{\bf p}_i^{\prime} = {\bf p}_{(i+k)\%K}
\end{equation}
and subsample the measurements $\{m_i\}$ to get corresponding responses
\begin{equation}
m_i^{\prime} = \left<m_{(i-1)n+\delta_-}, \cdots,m_{(i-1)n+\delta_+}\right>.
\end{equation}
Since the error caused by slight unsynchronization grows with the frequency increasing, so fractional synchronization is another advantage over the simple synchronization in \cite{welsh2013fast}.

At the final step, we can just exclude the synchronization patterns and retrieve the 2D scene from the synchronized measurements and illumination patterns. As for ghost imaging algorithms, there are mainly three variants---TGI\cite{bromberg2009ghost}, DGI\cite{ferri2010differential} and NGI\cite{sun2012normalized}---and each can be implemented in either iterative or batch manner. Among the three algorithms, we choose normalized ghost imaging due to its high robustness to external disturbances. To make full use of the high speed illumination, here we adopt the latter implementation by processing the entire data set in batch manner.
\begin{equation}
I = \left<\left(\frac{m_i^{\prime}}{\overline{{\bf p}_i^{\prime}}}-\frac{ \left<m_i^{\prime}\right> }{\overline{\left<{\bf p}_i^{\prime}\right>}}\right)\left({\bf p}_i^{\prime}-{\left<{\bf p}_i^{\prime}\right>}\right)\right>,
\end{equation}
in which $\overline{(\cdot)}$ computes the average of the elements in a pattern and $\left<\cdot\right>$ denotes temporal average.


Here we conduct a quantitative experiment on synthetic data to check the advantages of the proposed fractional self-synchronization strategy. We generate 10,000 100$\times$80 random binary patterns and calculate their inner products with a sharp 'text' image of the same resolution to get the latent correlated measurements. Then we denote the period of illumination to be $\tau$ and apply a left-half gaussian convolution with standard deviation being 0.24$\tau$ and 2.8$\times$$10^{-4}\tau$ successively to simulate the elapse in SLM and detector. As for system noise, gaussian white noise with standard deviation being 0.1 of the latent signal is added to obtain the final measurements from the detector.
To compare the results with and without fractional synchronization, we simulate the sampling process of acquisition card by sampling at 1$\times$ (non-fractional) and 100$\times$ (fractional) respectively.

\begin{figure}[h]
  \centering
  \includegraphics[width=\linewidth]{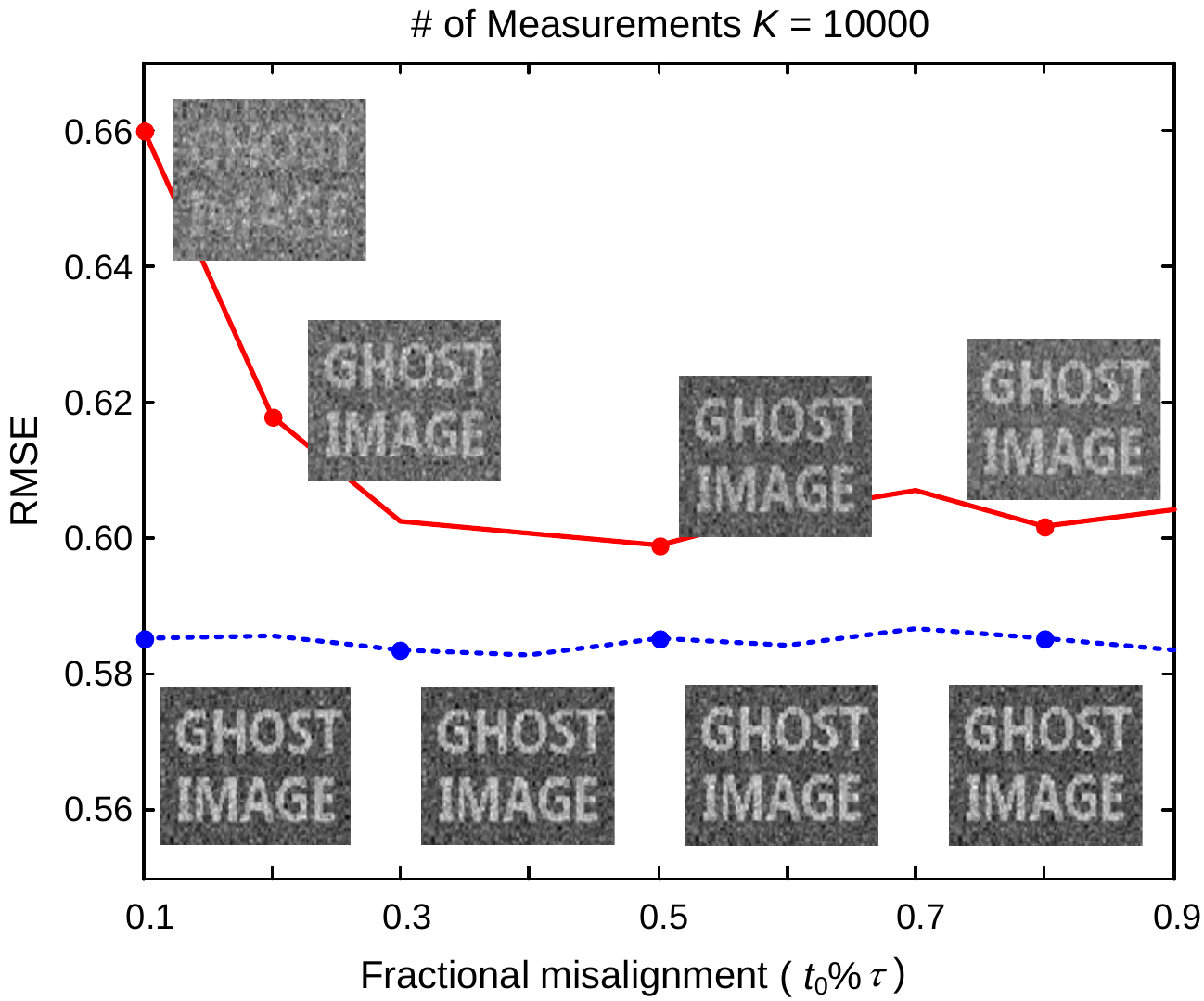}
  \caption{Quantitative comparison between results with and without fractional synchronization.}\label{fig:sim_alpha}
\end{figure}

Here we adopt root mean square error (RMSE) to compare the reconstruction qualities of two synchronizations at varying fractional misalignments using the data synthesized at different beginning instants $\{t_0$=$0.1\tau,0.2\tau,\cdots,0.9\tau\}$. 
The apparent smaller reconstruction error and better visual performance of fractional alignment in Fig.~\ref{fig:sim_alpha} clearly display its consistent performance advantage compared to the non-fractional synchronization.
As the starting instant varies, we can see that the fractionally synchronized reconstruction keeps at a high performance, while the result with only 1$\times$ sampling degenerates apparently in the transitional stage.
The reconstruction error for $t_0<0.3\tau$ is obviously larger than that of $t_0>0.2\tau$, and the reconstruction almost fails when $t_0=0.1\tau$.
The performance degeneration in the transitional stage of SLM is mainly due to the inaccurate match between illumination and measurement.
There also exists some  performance superiority of fractional alignment at the stabilized stage, and this advantage mainly comes from the noise suppression by averaging multiple valid (fine aligned) measurements. Therefore, we can maximize the sampling rate within the capability of detector and acquisition card to raise the final reconstruction quality.

\begin{figure}[t]
  \centering
  \vspace{2mm}
  \subfigure[]{
  \includegraphics[width=\linewidth]{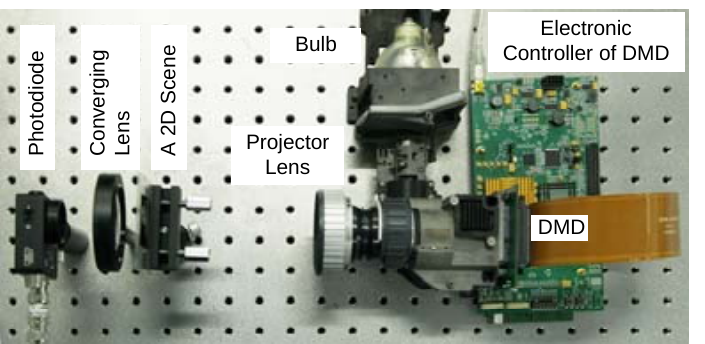}\label{fig:setup}}
  \subfigure[]{
  \includegraphics[width=0.223\textwidth]{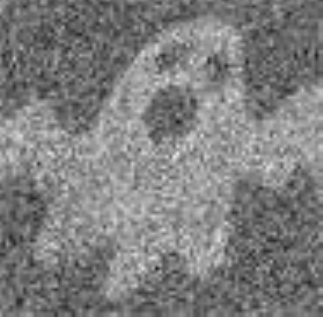}\label{fig:real1}}\hspace{2mm}
  \subfigure[]{
  \includegraphics[width=0.223\textwidth]{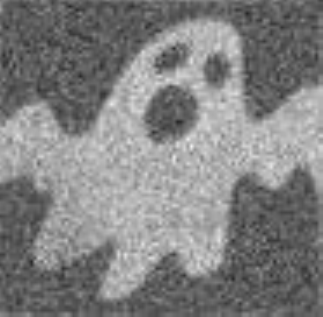}\label{fig:real2}}
  \caption{(a) The prototype system. (b)(c) Ghost imaging results of the proposed system and corresponding algorithm with 10,000 and 50,000 measurements respectively.}
\end{figure}

To validate the proposed scheme, we build a prototype system exhibited in Fig.~\ref{fig:setup}, whose light path is displayed in Fig.~\ref{fig:lightpath}. The system consists mainly two modules: the programmable illuminator is generated by a DMD (Texas Instrument DLP Discovery 4100, .7XGA) that can project binary patterns at 20kHz (the maximum 32kHz is so far inaccessible to our own SDK)  with 12$\mu$s 0/1 transition time, i.e., 24\% of the slot for each illumination pattern; a high speed single pixel detector is implemented by Thorlabs DET36A Silicon photodiode (350-1100 nm) with rising time being 14$ns$. Specifically, the illuminator is implemented by hacking an off-the-shelf DLP projector via replacing its built in DMD with our DLP $\textcircled{\small R}$ Discovery$^{\text{TM}}~4100$. The measurements encoding the scene information are collected by sampling the outputs of the detector using a 14bit acquisition board ART PCI8514, which digitalizes the analog signals and transports them to the computer for follow-up reconstruction. We keep the cooling system and power module of the original projector. In order to ensure high acquisition accuracy, we set the acquisition frequency at 100 times of the illumination, i.e., 2mHz, and each correlated measurement is averaged over 76 samples, which lie in the stabilized stage.

We use 200$\times$200 pixels out of the whole 1024$\times$768 pixels in the DMD. The synchronization patterns and random patterns are focused onto a 34mm$\times$34mm film, and the photons travel through the film are collected onto the detector by a converging lens ($\phi$=50.8mm, $f$=50.8mm). During capturing, we keep the detector working all through, and then project the preloaded binary patterns onto the film using DMD from an arbitrary instant. 
In sum, we need only a slot longer than 0.5 seconds to acquire 10,000 measurements and the imaging speed is remarkably higher than state-of-the-arts, which at most take around 1.4k measurements per second. The speed can be further raised by making full use the the 32kHz frequency of the DMD. One result reconstructed from 10,000 measurements acquired by our system is shown in Fig.~\ref{fig:real1}. Recall that the maximum number of projected binary patterns is currently limited by the internal memory size of our DMD controller board. The memory can be increased to shed a longer pattern sequence and improve the performance further but with only a linearly increasing acquisition time. To validate this potential performance of our approach, we load and project four more groups of random patterns (i.e., 50,000 in total) and concatenate the measurements for final reconstruction. One can see that the result is improved further, as displayed in Fig.~\ref{fig:real2}.

In summary, this letter proposes a self-synchronization scheme to perform fast ghost imaging using a pair of un-synchronized detector and illumination pattern, both of which work at constant frequencies.
The proposed approach largely addresses the synchronization challenge for fast ghost imaging, takes good advantage of the high frequency illumination/detector, and is flexible for replacing either illumination or detector module easily.

\newpage
\section*{Informational Fifth Page}

\end{document}